\documentclass[pra,aps,showpacs,twocolumn,superscriptaddress]{revtex4-1}
\usepackage{amsmath}
\usepackage{amssymb}
\usepackage{graphicx}
\usepackage{epstopdf}
\usepackage{easybmat}
\usepackage{color,soul}
\usepackage{times,txfonts}
\usepackage{ulem}
\usepackage{mathtools}
\newcommand{\ket}[1]{|#1\rangle}
\newcommand{\bra}[1]{\langle #1|}
\usepackage[bookmarks=false]{hyperref}
\hypersetup{colorlinks=true,citecolor=blue,linkcolor=blue,urlcolor=blue,pdfstartview=FitH,bookmarksopen=true}


\begin{document}

\title {Mitigation of systematic amplitude error in nonadiabatic holonomic operations}
\author{P. Z. Zhao}
\affiliation{Center for Quantum Technologies, National University of Singapore, Singapore 117543, Singapore}
\author{Jiangbin Gong}
\email{phygj@nus.edu.sg}
\affiliation{Center for Quantum Technologies, National University of Singapore, Singapore 117543, Singapore}
\affiliation{Department of Physics, National University of Singapore, Singapore 117551, Singapore}
\affiliation{Joint School of National University of Singapore and Tianjin University, International Campus of Tianjin University, Binhai New City, Fuzhou 350207, China}

\date{\today}

\begin{abstract}
Nonadiabatic holonomic operations are based on nonadiabatic non-Abelian geometric phases, hence possessing the inherent geometric features for robustness against control errors.
However, nonadiabatic holonomic operations are still sensitive to the systematic amplitude error induced by imperfect control of pulse timing or laser intensity. In this work, we present a scheme of nonadiabatic holonomic operations in order to mitigate the said systematic amplitude error. This is achieved by introducing a monitor qubit along with a conditional measurement on the monitor qubit that serves as an error correction device.  We shall show how to filter out the undesired effect of the systematic amplitude error, thereby improving the performance of nonadiabatic holonomic operations.
\end{abstract}

\maketitle

\section{Introduction}

Quantum operations are a basic element in many quantum information processing tasks, such as production of entanglement \cite{Hagley,Turchette}, quantum state population transfer \cite{Cirac,Bergmann}, quantum teleportation \cite{Bennett} and quantum computation \cite{Deutsch}.
Geometric phases are only dependent on the evolution path of the quantum system but independent of evolution details so that the quantum operation based on geometric phases possesses the inherent geometric features for robustness against control errors \cite{Chiara,Solonas,Zhu2005,Lipo,Filipp,Johansson,Berger,
Johansson}. The early schemes of geometric operations \cite{Zanardi,Jones,Duan} are based on Berry phases \cite{Berry} or adiabatic non-Abelian geometric phases \cite{Wilczek}.
However, the implementation of these schemes needs a long run time associated with adiabatic evolution \cite{Messiah,Tong}, which undoubtedly degrades its effectiveness due to the decoherence arising from the interaction between the quantum system and its environment.
To avoid this problem, nonadiabatic geometric operations \cite{WangXB,Zhu} based on nonadiabatic Abelian geometric phases \cite{Aharonov} and nonadiabatic holonomic operations \cite{Sjoqvist,Xu} based on nonadiabatic non-Abelian geometric phases \cite{Anandan} were proposed. The latter utilizes the so-called holonomic matrix as a building block of quantum operations and therefore possesses inherent geometric features for robustness against control errors.

The seminal scheme of nonadiabatic holonomic operations is performed with a resonant three-level system \cite{Sjoqvist,Xu}. This scheme needs to combine two $\pi$ rotations about different axes for realizing an arbitrary rotation operation. To simplify the realization, the single-shot scheme \cite{Xu2015,Sjoqvist2016} and sing-loop scheme \cite{Herterich} of nonadiabatic holonomic operations were put forward.
The two schemes enable an arbitrary rotation operation to be realized in a single-shot implementation, thereby reducing about half of the exposure time for nonadiabatic holonomic operations to error sources.
To further shorten the exposure time, a general approach of constructing Hamiltonians for nonadiabatic holonomic operations was put forward \cite{Zhao}.
Up to now, a number of physical implementations \cite{Spiegelberg,Liang,Zhang,Xue,Xue2016,Xue2017,Zhao2017,Hong,
Xue2018,Zhao2019,Xue2020,Xue2021,Xue2022,Liu,Zhao2023,Zhang2023} and experimental demonstrations \cite{Abdumalikov,Long,Duan2014,Sekiguchi,Yin2019,
Sun,Zhou,Camejo,Nagata} have been reported, greatly pushing forward the development of nonadiabatic holonomic quantum control.

For the preceding schemes of nonadiabatic holonomic operations, a common requirement is that the integration of laser intensity over a period of time should be equal to a constant number.
For example, in the seminal scheme \cite{Sjoqvist,Xu}, the holonomic operation $U=\boldsymbol{\mathrm{n}\cdot\sigma}$ is implemented using the Hamiltonian $H(t)=\Omega(t)(\ket{e}\bra{b}+\ket{b}\bra{e})$ with the requirement
$\int^{\tau}_{0}\Omega(t)dt=\pi$, where $\boldsymbol{\mathrm{n}}=(\sin\theta\cos\varphi,\sin\theta\sin\varphi,\cos\theta)$ is an arbitrary unit vector determining the orientation of a rotation axis,
$\boldsymbol{\sigma}=(\sigma_{x},\sigma_{y},\sigma_{z})$ is the standard Pauli operator, and $\ket{b}=\sin(\theta/2)\exp(-i\varphi)\ket{0}-\cos(\theta/2)\ket{1}$.
It is clear that the imperfect control of pulse timing $\tau$ or laser intensity $\Omega(t)$ will result in an inaccuracy of the integration $\int^{\tau}_{0}\Omega(t)dt$, namely, the systematic amplitude error.
This leads to the real output state deviating from the target output state, thereby becoming a crucial source of inaccurate quantum operations \cite{Jin,Jin2024}. In other words, due to the systematic amplitude error, the operations intended to be holonomic  are no longer purely geometrical in nature.
In actual experimental platforms such as nuclear magnetic resonance and trapped ions, the systematic amplitude error often occurs due to the technically  imperfect control on the prescribed amplitude and duration of a driving field.  In cases of nuclear magnetic resonance experiments, the radio-frequency field inhomogeneity and imperfect pulse length calibration can lead to the systematic amplitude error \cite{Vandersypen}. In  trapped-ion experiments, the temperature changes and voltage fluctuations for the trap electrodes can lead to a drift of the vibrational quantum numbers and hence an inaccurate Rabi oscillation for the sideband laser pulse \cite{Monz,Timoney}.

To date, there are two approaches to mitigating the systematic amplitude error in nonadiabatic holonomic operations. One approach is the composite nonadiabatic holonomic operations \cite{Xu2017}. This approach adopts the basic ideal of composite pulse technology and thus can effectively suppress the systematic amplitude error. However, it needs four elementary operations to implement an arbitrary rotational operation. The increased number of unitary operations extends the total evolution time, consequently amplifying the impact of environment-induced decoherence.
The other approach exploits environment-assisted nonadiabatic holonomic operations \cite{Ramberg}. This approach needs to engineer the environment of a quantum system to minimize the systematic amplitude error. However, in many situations the environment might not be easily controllable and hence it can be  a challenge to engineer the actual environment for a quantum system of interest.

In this work, we put forward a postselection-based scheme of nonadiabatic holonomic operations in order to mitigate the systematic amplitude error.  We propose to introduce a monitor qubit and then utilize a conditional measurement on the monitor qubit to filter out the unwanted systematic amplitude error occurring on the computational qubit. Clearly then, the monitor qubit introduced here serves as an error-correction device, through which we can improve the fidelity of our operations.
Our scheme thus represents a measurement-assisted approach towards more accurate nonadiabatic holonomic quantum control.

\section{Scheme}

Consider a quantum system depicted by a Hilbert space $\mathcal{H}$. This quantum system is comprised of two subsystems, named principal subsystem $\mathcal{H}_{P}$ and monitoring subsystem $\mathcal{H}_{M}$.
The principal subsystem $\mathcal{H}_{P}$ is partitioned into an $L$-dimensional data subspace  $\mathcal{H}^{a}_{P}(t)=\mathrm{Span}\{\ket{\phi_{k}(t)}\}^{L}_{k=1}$ and a one-dimensional auxiliary subspace $\mathcal{H}^{b}_{P}(t)=\mathrm{Span}\{\ket{\phi_{b}(t)}\}$,
where $t$ is the time variable, and $\ket{\phi_{k}(t)}$ and $\ket{\phi_{b}(t)}$ are the time-dependent orthonormal basis in $\mathcal{H}_{P}$.
The initial subspace of $\mathcal{H}^{a}_{P}(t)$ is used as the computational subspace.
A computational qubit is generally represented by a two-level system, hence it is reasonable to take $L=2$. In such a case, the data subspace is reduced to a two-dimensional subspace $\mathcal{H}^{a}_{P}(t)=\mathrm{Span}\{\ket{\phi_{1}(t)},\ket{\phi_{2}(t)}\}$ with the feature
$\mathcal{H}^{a}_{P}(0)=\mathrm{Span}\{\ket{\phi_{1}(0)},\ket{\phi_{2}(0)}\}
=\mathrm{Span}\{\ket{0},\ket{1}\}$.
The monitoring subsystem $\mathcal{H}_{M}$ is partitioned into two one-dimensional subspaces
$\mathcal{H}^{a}_{M}(t)=\mathrm{Span}\{\ket{a(t)}\}$ and $\mathcal{H}^{b}_{M}(t)=\mathrm{Span}\{\ket{b(t)}\}$, where the time-dependent orthonormal basis is set to cyclic vectors such that
$\ket{a(\tau)}=\ket{a(0)}\equiv\ket{a}$ and $\ket{b(\tau)}=\ket{b(0)}\equiv\ket{b}$ with $\tau$ being the total time of a quantum operation.

\begin{figure}[t]
  \includegraphics[scale=0.25]{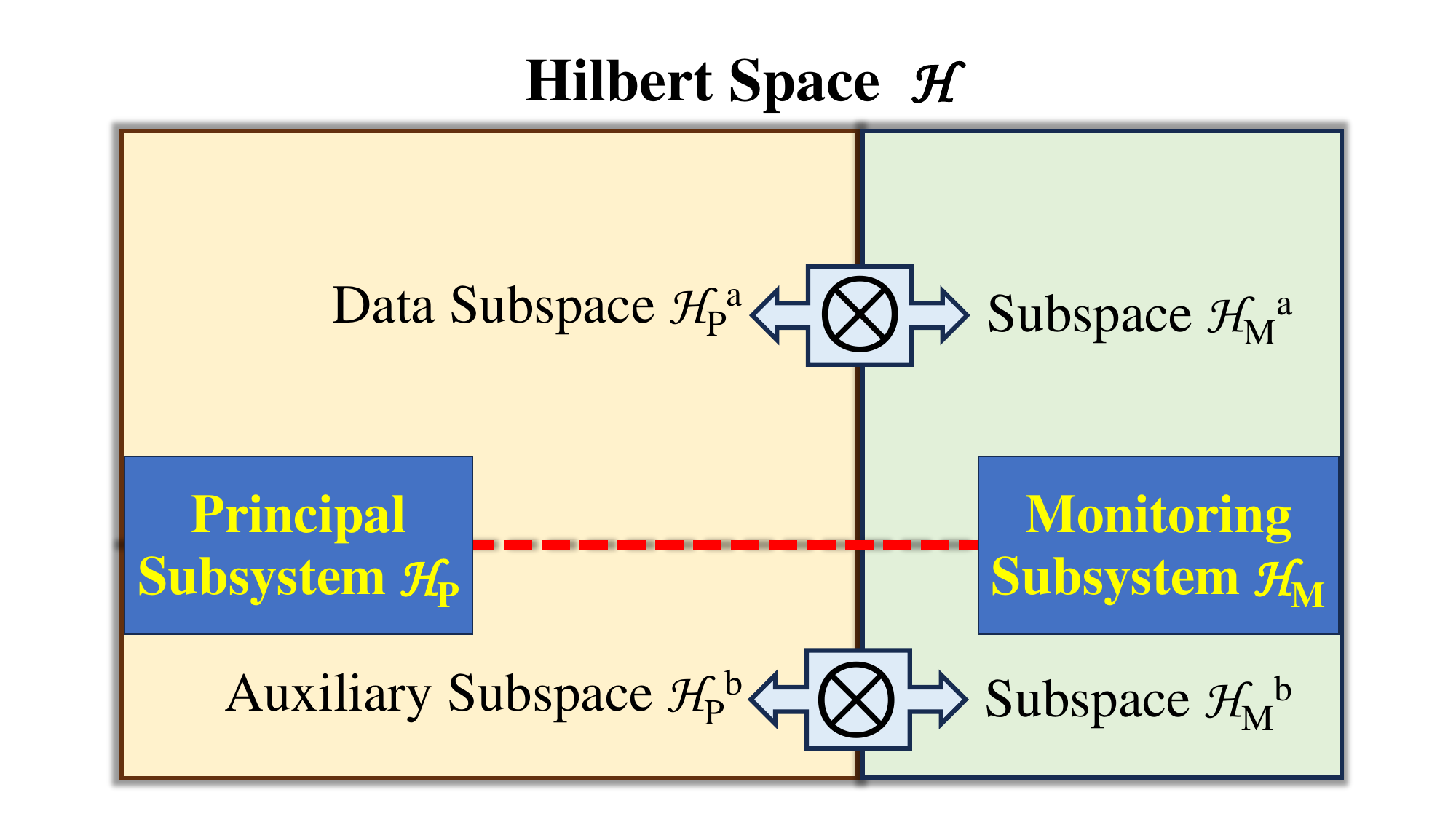}
  \caption{The decomposition construction of the Hilbert space.}
  \label{Fig1}
\end{figure}

The starting point of our scheme is to require the Hilbert space to possess the following mathematical structure
\begin{align}\label{eq}
\mathcal{H}=[\mathcal{H}^{a}_{P}(t)\otimes\mathcal{H}^{a}_{M}(t)]
\oplus[\mathcal{H}^{b}_{P}(t)\otimes\mathcal{H}^{b}_{M}(t)],
\end{align}
schematically shown in Fig.~\ref{Fig1}.
In light of this requirement, $\ket{\phi_{1}(t)}\otimes\ket{a(t)}$, $\ket{\phi_{2}(t)}\otimes\ket{a(t)}$, and $\ket{\phi_{b}(t)}\otimes\ket{b(t)}$ consist of a set of orthonormal basis vectors in the Hilbert space $\mathcal{H}$.
It is clear that this requirement establishes a connection between the data subspace $\mathcal{H}^{a}_{P}(t)$ and the monitoring subspace $\mathcal{H}^{a}_{M}(t)$, and simultaneously links the auxiliary subspace $\mathcal{H}^{b}_{P}(t)$ to the monitoring subspace $\mathcal{H}^{b}_{M}(t)$.
As a consequence, it is now possible to mitigate the systematic amplitude error occurring on the quantum operation for the computational subspace by performing a conditional measurement or a postselection on the monitoring subsystem.

For our purpose, we suppose that $\ket{\phi_{1}(t)}\otimes\ket{a(t)}$, $\ket{\phi_{2}(t)}\otimes\ket{a(t)}$, and $\ket{\phi_{b}(t)}\otimes\ket{b(t)}$ are the solutions of the Schr\"{o}dinger equation $i\ket{\dot{\psi}(t)}=H(t)\ket{\psi(t)}$, where $H(t)$ is the driving Hamiltonian governing the time evolution of the quantum system.
As an example, we take the driving Hamiltonian as
\begin{align}\label{eq0}
H(t)=\Omega(t)\ket{\phi_{b}(0)}\bra{\phi_{2}(0)}\otimes\ket{b}\bra{a}+\mathrm{H.c.},
\end{align}
where $\Omega(t)$ is a time-dependent real parameter, and $\mathrm{H.c.}$ represents the Hermitian conjugate term.
For this Hamiltonian, the basis state $\ket{\phi_{1}(0)}\otimes\ket{a}$ is a dark state, which can be seen from the fact that $H(t)\ket{\phi_{1}(0)}\otimes\ket{a}=0$.
The evolution operator corresponding to the Hamiltonian reads
\begin{align}
\mathcal{U}(t)=&[\ket{\phi_{1}(0)}\bra{\phi_{1}(0)}\otimes\ket{a}\bra{a}
+\cos\omega(t)[\ket{\phi_{2}(0)}\bra{\phi_{2}(0)}
\notag\\
&\otimes\ket{a}\bra{a}
+\ket{\phi_{b}(0)}\bra{\phi_{b}(0)}\otimes\ket{b}\bra{b}]-i\sin\omega(t)
\notag\\
&\times[\ket{\phi_{b}(0)}\bra{\phi_{2}(0)}\otimes\ket{b}\bra{a}+\mathrm{H.c.}]
\end{align}
with $\omega(t)=\int^{t}_{0}\Omega(t^{\prime})dt^{\prime}$.
If we require
\begin{align}\label{eq1}
\int^{\tau}_{0}\Omega(t)dt=\pi,
\end{align}
the above defined evolution operator then yields
\begin{align}\label{eq2}
\mathcal{U}(\tau)=&\left[\ket{\phi_{1}(0)}\bra{\phi_{1}(0)}-\ket{\phi_{2}(0)}\bra{\phi_{2}(0)}\right]
\otimes\ket{a}\bra{a}\notag\\
&-\ket{\phi_{b}(0)}\bra{\phi_{b}(0)}\otimes\ket{b}\bra{b}.
\end{align}
Considering that the initial state in the principal subsystem resides in the computational subspace spanned by the basis $\ket{\phi_{1}(0)}$ and $\ket{\phi_{2}(0)}$, the unitary operation is actually equivalent to $\mathcal{U}(\tau)=[\ket{\phi_{1}(0)}\bra{\phi_{1}(0)}-
\ket{\phi_{2}(0)}\bra{\phi_{2}(0)}]\otimes\ket{a}\bra{a}$.
Thereafter, the target operation can be obtained by tracing out $\ket{a}\bra{a}$, which yields
\begin{align}\label{eqq}
U=\ket{\phi_{1}(0)}\bra{\phi_{1}(0)}-\ket{\phi_{2}(0)}\bra{\phi_{2}(0)}.
\end{align}
It defines a rotation operation about the axis determined by $\{\ket{\phi_{1}(0)},\ket{\phi_{2}(0)}\}$ with an angle $\pi$.
Especially, when $\ket{\phi_{1}(0)}$ and $\ket{\phi_{1}(0)}$ are set to
$\ket{\phi_{1}(0)}=\cos(\theta/2)\ket{0}+\sin(\theta/2\exp(i\varphi)\ket{1}$ and $\ket{\phi_{2}(0)}=\sin(\theta/2)\exp(-i\varphi)\ket{0}-\cos(\theta/2)\ket{1}$, i.e., the eigenstates of $\boldsymbol{\mathrm{n}\cdot\sigma}$,
the rotation axis along an arbitrary direction can be implemented.

Before proceeding further, we briefly demonstrate that the evolution operator in our scheme is a nonadiabatic holonomic operation. Nonadiabatic holonomic transformation arises from the time evolution of a quantum system with a subspace, for example the subspace $\mathcal{H}^{a}_{P}(t)\otimes\mathcal{H}^{a}_{M}(t)$ in the Hilbert space $\mathcal{H}$ of our scheme,
satisfying the cyclic evolution condition
\begin{align}
&\sum^{L}_{k=1}\ket{\phi_{k}(\tau)}\bra{\phi_{k}(\tau)}\otimes\ket{a(\tau)}\bra{a(\tau)}
\notag\\
=&\sum^{L}_{k=1}\ket{\phi_{k}(0)}\bra{\phi_{k}(0)}\otimes\ket{a(0)}\bra{a(0)}
\end{align}
and the parallel transport condition \cite{Sjoqvist,Xu}
\begin{align}
\bra{\phi_{k}(t)}\otimes\bra{a(t)}H(t)\ket{a(t)}\otimes\ket{\phi_{l}(t)}=0.
\end{align}
The first condition guarantees that a quantum state in the desired subspace returns to the initial subspace and the second condition  ensures that the unitary operator is purely geometric within the subspace.
From Eq.~(\ref{eq2}), we can readily verify that the cyclic evolution condition is satisfied.
Furthermore,   using the commutation relation $[H(t),\mathcal{U}(t)]=0$, one can confirm that $\bra{\phi_{k}(t)}\otimes\bra{a(t)}H(t)\ket{a(t)}\otimes\ket{\phi_{l}(t)}
=\bra{\phi_{k}(0)}\otimes\bra{a(0)}\mathcal{U}^{\dagger}(t)H(t)\mathcal{U}(t)
\ket{a(0)}\otimes\ket{\phi_{l}(0)}
=\bra{\phi_{k}(0)}\otimes\bra{a(0)}H(t)\ket{a(0)}\otimes\ket{\phi_{l}(0)}=0$, i.e, the parallel transport condition is also satisfied.
The unitary time evolution considered above is hence a nonadiabatic holonomic operation.

The above discussion is the ideal case without any operational errors.  In practice, it is difficult to execute  perfect control on the quantum system. The imperfect control may lead to the quantum system over-evolving or under-evolving during the time evolution, resulting in the output state leaking into the entire Hilbert space (hence no longer purely geometrical).   A typical control imperfection that induces leakage is the systematic amplitude error, which occurs in such a way that
\begin{align}\label{eq3}
\int^{\tau}_{0}\Omega(t)dt=\pi\rightarrow(1+\epsilon)\pi
\end{align}
owing to the imperfect controls of evolution time $\tau\rightarrow(1+\epsilon)\tau$ or amplitude parameter $\Omega(t)\rightarrow(1+\epsilon)\Omega(t)$.
In this case, the resulting unitary operator is found to be
\begin{align}
\mathcal{U}_{\epsilon}(\tau)=&\left[\ket{\phi_{1}(0)}\bra{\phi_{1}(0)}
-\cos(\epsilon\pi)\ket{\phi_{2}(0)}\bra{\phi_{2}(0)}\right]
\otimes\ket{a}\bra{a}
\notag\\
&-\cos(\epsilon\pi)\ket{\phi_{b}(0)}\bra{\phi_{b}(0)}\otimes\ket{b}\bra{b}
\notag\\
&+i\sin(\epsilon\pi)\big[\ket{\phi_{2}(0)}\bra{\phi_{b}(0)}\otimes\ket{a}\bra{b}
\notag\\
&+\ket{\phi_{b}(0)}\bra{\phi_{2}(0)}\otimes\ket{b}\bra{a}\big].
\end{align}
Still assuming that the initial state of the principal subsystem
resides in the computational subspace, we always consider the initial state
$\ket{\psi(0)}=[c_{1}\ket{\phi_{1}(0)}+c_{2}\ket{\phi_{2}(0)}]\otimes\ket{a}$, where $|c_{1}|^{2}+|c_{2}|^{2}=1$.  To see clearly what the systematic amplitude error may lead to, let us recall again that in the ideal case, the final state is given by
$\ket{\psi(\tau)}=[c_{1}\ket{\phi_{1}(0)}-c_{2}\ket{\phi_{2}(0)}]\otimes\ket{a}$ and
the target output state after performing a partial trace yields
\begin{align}\label{eq4}
\ket{\phi(\tau)}=c_{1}\ket{\phi_{1}(0)}-c_{2}\ket{\phi_{2}(0)}.
\end{align}
By contrast, in a nonideal case with the systematic amplitude error, the evolution state under the action of $\mathcal{U}_{\epsilon}(\tau)$ is in turn given by
\begin{align}
\ket{\psi_{\epsilon}(\tau)}=&[c_{1}\ket{\phi_{1}(0)}-c_{2}\cos(\epsilon\pi)\ket{\phi_{2}(0)}]\otimes\ket{a}
\notag\\
&+ic_{2}\sin(\epsilon\pi)\ket{\phi_{b}(0)}\otimes\ket{b}.
\end{align}
Obviously, the systematic amplitude error leads to the evolution state leaking into the entire Hilbert space.
However, our scheme is designed in such a way that we are allowed to monitor the impact of the systematic amplitude error on the monitor qubit and hence acquire indirectly some information about the quality of the operation. Specifically, at the end of the time evolution, we can always perform a projective measurement on the monitoring subsystem to project the state back to the initial computational subspace.
This projective measurement yields
\begin{align}\label{eq5}
\ket{\phi_{\epsilon}(\tau)}=\frac{c_{1}\ket{\phi_{1}(0)}
-c_{2}\cos(\epsilon\pi)\ket{\phi_{2}(0)}}{\sqrt{|c_{1}|^{2}+|c_{2}|^{2}\cos^{2}(\epsilon\pi)}}
\end{align}
when collapsing the monitor into the basis vector $\ket{a}$, and $\ket{\bar{\phi}_{\epsilon}(\tau)}=-\ket{\phi_{b}(0)}$ when collapsing the monitor into the basis vector $\ket{b}$.
As a consequence, conditional on $\ket{a}$ being detected, we claim that the output state is obtained, reading $\ket{\phi_{\epsilon}(\tau)}$.
As indicated from Eq.~(12), the success probability of this postselection is $|c_{1}|^{2}+|c_{2}|^{2}\cos^{2}(\epsilon\pi)$. For a small systematic amplitude error (i.e., small $\epsilon$), the success probability is approximately equal to $1-|\pi{c}_2|^2\epsilon^2+|\pi^2{c}_2|^2\epsilon^4/4$, showing that the robust output state can be obtained with a rather high success probability.

The above-assumed Hilbert space structure to implement our scheme is relevant to actual physical systems. For example, the principal subsystem can be taken as a three-level material qutrit while the monitor subsystem can be taken as a photon qubit in cavity quantum electrodynamics \cite{Pellizzari}, a vibrational mode in trapped ions \cite{Cirac1995}, an oscillator mode in superconducting quantum circuits \cite{Blais,Blais2007} and so on. The Hamiltonian in Eq.~(\ref{eq0}) is then commonly expressed as $H(t)=\Omega_{0}(t)\ket{e}\bra{0}\otimes\ket{0}\bra{1}
+\Omega_{1}(t)\ket{e}\bra{0}\otimes\ket{0}\bra{1}+\mathrm{H.c.}$,
where the first term of the tensor product denotes the principal qutrit and the second term denotes the monitor qubit. Here, we set the parameters to be $\Omega_{0}(t)=\Omega(t)\sin(\theta/2)\exp(i\varphi)$ and $\Omega_{1}=-\Omega(t)\cos(\theta)$, and then we have
$\ket{\phi_{1}(0)}=\cos(\theta/2)\ket{0}+\sin(\theta/2\exp(i\varphi)\ket{1}$,
$\ket{\phi_{2}(0)}=\sin(\theta/2)\exp(-i\varphi)\ket{0}-\cos(\theta/2)\ket{1}$,
and $\ket{\phi_{b}(0)}=\ket{e}$. The driving can be facilitated by coupling the atomic states to the quantized cavity mode through Jaynes-Cummings interactions in cavity quantum electrodynamics, by coupling the internal states to the motional levels through the sideband transition in trapped ions, or by coupling the transmon to a superconducting transmission line resonator in superconducting quantum circuits. All these options indicate that our scheme is already possible based on  hardware available today.

\section{Performance}

Let us now compare what our scheme can achieve versus what happens we do not resort to the monitoring subsystem.
We consider the seminal scheme of nonadiabatic holonomic operations \cite{Sjoqvist,Xu}. Therein, the driving Hamiltonian is taken as $H^{\prime}(t)=\Omega(t)(\ket{\phi_{b}(0)}\bra{\phi_{2}(0)}+\mathrm{H.c.})$ [corresponding to the first term of the tensor product in Eq.~(\ref{eq0})] with the requirement $\int^{\tau}_{0}\Omega(t)dt=\pi$. The resulting evolution operator acting on the computational subspace yields $U^{\prime}=\ket{\phi_{1}(0)}\bra{\phi_{1}(0)}-\ket{\phi_{2}(0)}\bra{\phi_{2}(0)}$, like the form in Eq.~(\ref{eqq}).
If there is a systematic amplitude error such as that in Eq.~(\ref{eq3}), the time evolution operator is given by
\begin{align}
U^{\prime}_{\epsilon}(\tau)=&\ket{\phi_{1}(0)}\bra{\phi_{1}(0)}
-\cos(\epsilon\pi)\ket{\phi_{2}(0)}\bra{\phi_{2}(0)}
\notag\\
&-\cos(\epsilon\pi)\ket{\phi_{b}(0)}\bra{\phi_{b}(0)}+i\sin(\epsilon\pi)
\notag\\
&\times\left[\ket{\phi_{2}(0)}\bra{\phi_{b}(0)}+\ket{\phi_{b}(0)}\bra{\phi_{2}(0)}\right].
\end{align}
For the same initial input state $\ket{\phi(0)}=c_{1}\ket{\phi_{1}(0)}+c_{2}\ket{\phi_{2}(0)}$ residing in the computational subspace, the output state under the action of $U^{\prime}_{\epsilon}$ then reads
\begin{align}\label{eq6}
\ket{\phi^{\prime}_{\epsilon}(\tau)}=&c_{1}\ket{\phi_{1}(0)}-c_{2}\cos(\epsilon\pi)\ket{\phi_{2}(0)}
\notag\\
&+ic_{2}\sin(\epsilon\pi)\ket{\phi_{b}(0)}.
\end{align}
As seen above, the systematic amplitude error causes the time evolving state to leak out of the computational subspace.  However, in the plain version, there is no extra qubit tagging the unwanted amplitude on the state $\ket{\phi_{b}(0)}$.

\begin{figure}[t]
  \includegraphics[scale=0.58]{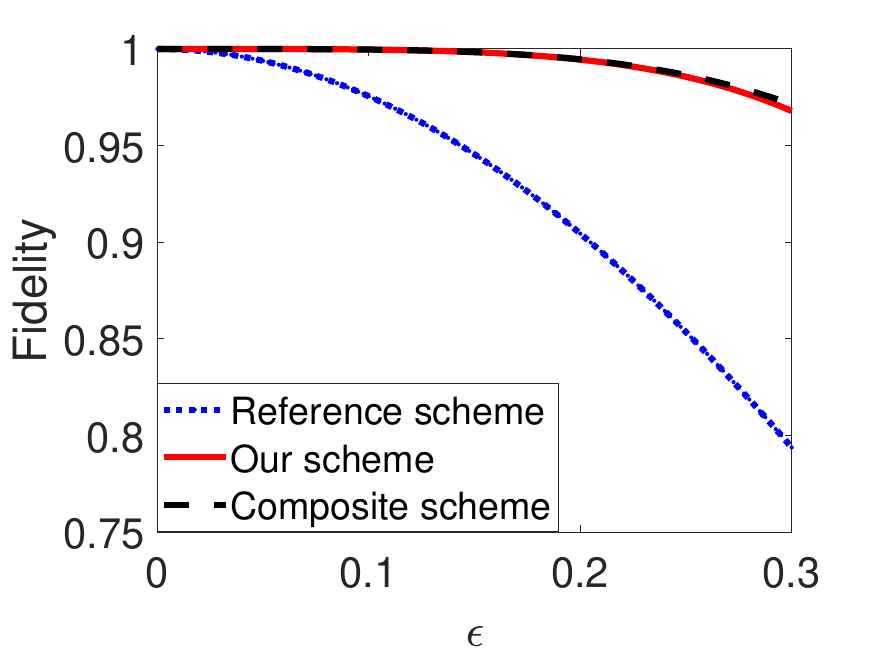}
  \caption{The fidelities of our scheme, the reference scheme, and the composite scheme.  Note, however, that a similar performance of the composite pulse scheme as compared with our scheme proposed in this work is obtained without accounting for any decoherence effects.   }
  \label{Fig2}
\end{figure}

To demonstrate the improvement of our nonadiabatic holonomic operation with a monitoring qubit, we compare fidelities $F=|\bra{\phi_{\epsilon}(\tau)}\phi(\tau)\rangle|$ of our scheme with the reference plain scheme described above, where $\ket{\phi_{\epsilon}(\tau)}$ is the erroneous output state and $\ket{\phi(\tau)}$ is the target output state.
A straightforward calculation based on Eqs.~(\ref{eq4}) and (\ref{eq5}) yields that in our scheme, the fidelity is given by
\begin{align}
F=\frac{|c_{1}|^{2}+|c_{2}|^{2}\cos(\epsilon\pi)}{\sqrt{|c_{1}|^{2}+|c_{2}|^{2}\cos^{2}(\epsilon\pi)}}.
\end{align}
In contrast, the fidelity of the reference scheme is obtained by combining Eqs.~(\ref{eq4}) and (\ref{eq6}) as
\begin{align}
F^{\prime}=|c_{1}|^{2}+|c_{2}|^{2}\cos(\epsilon\pi).
\end{align}
It is obvious that $F>F^{\prime}$.  Evidently then, our scheme improves the fidelity of nonadiabatic holonomic  operations on the computational space by introducing a conditional measurement on the monitor qubit. To elucidate the advantages of our approach, we further plot the fidelities $F$ (the red line) and $F^{\prime}$ (the blue line) vs the error ratio $\epsilon$ in Fig.~\ref{Fig2}, setting $c_{1}=c_{2}=1/\sqrt{2}$ for convenience.
The result shows that our scheme maintains high fidelity over the range $\epsilon\in[0,0.3]$ compared with the reference scheme. This indicates that our scheme indeed improves considerably the fidelity of nonadiabatic holonomic operations.
It is worth emphasizing that even for a large value $\epsilon=0.3$,
in the sense that the systematic amplitude error occurs in such a way that
$\int^{\tau}_{0}\Omega(t)dt=\pi\rightarrow1.3\pi$, the fidelity of our scheme still exceeds $95\%$ but the fidelity of the reference plain scheme is lower than $80\%$. This example indicates that the nonadiabatic holonomic operation in our scheme behaves well while the nonadiabatic holonomic operation in the reference scheme is strongly deteriorated by the systematic amplitude error.

Note that the previously mentioned composite nonadiabatic holonomic operations can also effectively mitigate the systematic amplitude error but extending the total evolution time \cite{Xu2017}.
Here, we illustrate these two points. In the composite nonadiabatic holonomic scheme, we need to sequentially applying the driving Hamiltonians $H_{1}=\Omega(t)(\ket{\phi_{b}(0)}\bra{\phi_{2}(0)}+\mathrm{H.c.})$
and $H_{2}=\Omega(t)(-i\ket{\phi_{b}(0)}\bra{\phi_{2}(0)}+\mathrm{H.c.})$ to the quantum system
with the requirement
$\int^{2\tau}_{\tau}\Omega(t)dt=\int^{\tau}_{0}\Omega(t)dt=\pi/2$.
The resulting operator is an elementary gate $U_{e}=\ket{\phi_{1}(0)}\bra{\phi_{1}(0)}+i\ket{\phi_{2}(0)}\bra{\phi_{2}(0)}$.
To realize the target evolution in Eq.~(\ref{eqq}), we need to combine two elementary gates such that  $U=U^{2}_{e}=\ket{\phi_{1}(0)}\bra{\phi_{1}(0)}-\ket{\phi_{2}(0)}\bra{\phi_{2}(0)}$.
Clearly then, this scheme increases the number of unitary operations and hence extends the operation duration, consequently amplifying the potential impact of environment-induced decoherence.
If there is a systematic amplitude error, the time evolution operator then yields
\begin{align}
U^{\prime}_{\epsilon}(\tau)=&\ket{\phi_{1}(0)}\bra{\phi_{1}(0)}
-\left[\cos(\epsilon\pi)+\frac{1}{\sqrt{2}}
\sin^{2}(\epsilon\pi)e^{\frac{i\pi}{4}}\right]
\notag\\
&\times\ket{\phi_{2}(0)}\bra{\phi_{2}(0)}
-\left[\cos(\epsilon\pi)+\frac{1}{\sqrt{2}}
\sin^{2}(\epsilon\pi)e^{-\frac{i\pi}{4}}\right]\notag\\
&\times\ket{\phi_{b}(0)}\bra{\phi_{b}(0)}.
\end{align}
For the input state $\ket{\phi(0)}=c_{1}\ket{\phi_{1}(0)}+c_{2}\ket{\phi_{2}(0)}$, the resulting output state is given by
\begin{align}
\ket{\phi^{\prime}_{\epsilon}(\tau)}=&c_{1}\ket{\phi_{1}(0)}-c_{2}\cos(\epsilon\pi)\ket{\phi_{2}(0)}
\notag\\
&-\frac{c_{2}}{\sqrt{2}}\sin^{2}(\epsilon\pi)e^{\frac{i\pi}{4}}\ket{\phi_{2}(0)}.
\end{align}
The fidelity between this erroneous state and the target output state in Eq.~(\ref{eq4}) is then calculated as
\begin{align}
\mathcal{F}^{\prime}=&\bigg\{\left[|c_{1}|^{2}+|c_{2}|^{2}\cos(\epsilon\pi)\right]^2
+\frac{|c_{2}|^{4}}{2}\sin^{4}(\epsilon\pi)
\notag\\
&+\left[|c_{1}|^{2}+|c_{2}|^{2}\cos(\epsilon\pi)\right]|c_{2}|^{2}\sin^{2}(\epsilon\pi)\bigg\}^{1/2}.
\end{align}
To elucidate the mitigation of the systematic amplitude errors, we additionally plot the fidelity $\mathcal{F}^{\prime}$ (black line) versus the error ratio $\epsilon$ in Fig.~\ref{Fig2}.
It shows that the composite nonadiabatic holonomic operations has a similar robustness to our scheme against systematic amplitude errors. This in turn indicates that our scheme is indeed robust against systematic amplitude errors while consuming much less evolution time compared with the composite  pulse scheme.

Finally, we would like to remark that our scheme can be generalized to other models of nonadiabatic holonomic operations \cite{Xu2015,Sjoqvist2016,Herterich,Zhao}. To illustrate this point, we take the model of the single-loop scheme \cite{Herterich} as another example. To that end, we divide the whole evolution into two equal-interval time evolutions with the Hamiltonians in the first interval $[0,\tau_{1})$ reading $H_{1}(t)=\Omega(t)(\ket{\phi_{b}(0)}\bra{\phi_{2}(0)}\otimes\ket{b}\bra{a}+\mathrm{H.c.})$ and in the second interval $[\tau_{1},\tau]$ reading $H_{2}(t)=\Omega(t)[\exp(-i\phi)\ket{\phi_{b}(0)}\bra{\phi_{2}(0)}\otimes\ket{b}\bra{a}+\mathrm{H.c.}]$, where $\Omega(t)$ is a time-dependent real parameter and $\phi$ is an arbitrary phase.
For these two Hamiltonians, $\ket{\phi_{1}(0)}$ is still a dark state. With the requirement $\int^{\tau_{1}}_{0}\Omega(t)dt=\int^{\tau}_{\tau_{1}}\Omega(t)dt=\pi/2$, we have
\begin{align}
\mathcal{U}(\tau)=&\left[\ket{\phi_{1}(0)}\bra{\phi_{1}(0)}-e^{i\phi}\ket{\phi_{2}(0)}\bra{\phi_{2}(0)}\right]
\otimes\ket{a}\bra{a}\notag\\
&-e^{-i\phi}\ket{\phi_{b}(0)}\bra{\phi_{b}(0)}\otimes\ket{b}\bra{b}.
\end{align}
Because the initial state in the principal subsystem is prepared in the computational subspace, the unitary operation is equivalent to $\mathcal{U}(\tau)=[\ket{\phi_{1}(0)}\bra{\phi_{1}(0)}-\exp(i\phi)\ket{\phi_{2}(0)}\bra{\phi_{2}(0)}
\otimes\ket{a}\bra{a}$.
Thereafter, we trace out the monitor qubit and then the target operation can be obtained as
\begin{align}
U=\ket{\phi_{1}(0)}\bra{\phi_{1}(0)}-e^{i\phi}\ket{\phi_{2}(0)}\bra{\phi_{2}(0)}.
\end{align}
Ignoring an unimportant global phase, it is just a rotation about an arbitrary axis with an arbitrary angle.
Without loss of generality, we consider an input state as $\ket{\psi(0)}=[c_{1}\ket{\phi_{1}(0)}+c_{2}\ket{\phi_{2}(0)}]\otimes\ket{a}$.
In the ideal case, the target output state after performing measurement yields
\begin{align}
\ket{\phi(\tau)}=c_{1}\ket{\phi_{1}(0)}-c_{2}e^{i\phi}\ket{\phi_{2}(0)}.
\end{align}
If the systematic amplitude error occurs in such a way that
the integration varies from $\pi/2$ to $(1+\epsilon)\pi/2$, the unitary operator then becomes
\begin{align}
\mathcal{U}_{\epsilon}(\tau)=&\bigg[\ket{\phi_{1}(0)}\bra{\phi_{1}(0)}
+\left(\sin^{2}\frac{\pi\epsilon}{2}
-\cos^{2}\frac{\pi\epsilon}{2}e^{i\phi}\right)
\notag\\
&\times\ket{\phi_{2}(0)}\bra{\phi_{2}(0)}\bigg]
\otimes\ket{a}\bra{a}
+\left(\sin^{2}\frac{\pi\epsilon}{2}
-\cos^{2}\frac{\pi\epsilon}{2}e^{-i\phi}\right)
\notag\\
&\times\ket{\phi_{b}(0)}\bra{\phi_{b}(0)}\otimes\ket{b}\bra{b}
+i\sin(\pi\epsilon)\cos\frac{\phi}{2}e^{\frac{i\phi}{2}}\notag\\
&\times\ket{\phi_{2}(0)}\bra{\phi_{b}(0)}\otimes\ket{a}\bra{b}
+i\sin(\pi\epsilon)\cos\frac{\phi}{2}e^{-\frac{i\phi}{2}}\notag\\
&\times\ket{\phi_{b}(0)}\bra{\phi_{2}(0)}\otimes\ket{b}\bra{a}.
\end{align}
The evolution state is consequently given by
\begin{align}
\ket{\psi_{\epsilon}(\tau)}=&\left[c_{1}\ket{\phi_{1}(0)}+c_{2}\left(\sin^{2}\frac{\pi\epsilon}{2}
-\cos^{2}\frac{\pi\epsilon}{2}e^{i\phi}\right)\ket{\phi_{2}(0)}\right]\otimes\ket{a}
\notag\\
&+i\sin(\pi\epsilon)\cos\frac{\phi}{2}e^{-\frac{i\phi}{2}}
\ket{\phi_{b}(0)}\otimes\ket{b}.
\end{align}
Afterwards, we perform a measurement on the monitor qubit. Then, we have the output state
\begin{align}
\ket{\phi_{\epsilon}(\tau)}=\frac{c_{1}\ket{\phi_{1}(0)}+c_{2}\left[\sin^{2}(\pi\epsilon/2)
-\cos^{2}(\pi\epsilon/2)e^{i\phi}\right]\ket{\phi_{2}(0)}}
{\sqrt{|c_{1}|^{2}+|c_{2}|^{2}\left[1-\sin^{2}(\epsilon\pi)\cos^{2}(\phi/2)\right]}}
\end{align}
conditional on the basis vector $\ket{a}$ clicking.
If we do not resort to the monitor qubit, the output state instead yields
\begin{align}
\ket{\phi^{\prime}_{\epsilon}(\tau)}=&c_{1}\ket{\phi_{1}(0)}+c_{2}\left(\sin^{2}\frac{\pi\epsilon}{2}
-\cos^{2}\frac{\pi\epsilon}{2}e^{i\phi}\right)\ket{\phi_{2}(0)}
\notag\\
&+i\sin(\pi\epsilon)\cos\frac{\phi}{2}e^{-\frac{i\phi}{2}}
\ket{\phi_{b}(0)}.
\end{align}
Using the same method to calculate the fidelity as before, we can also conclude that the output state with our scheme is closer to the ideal case than that without introducing any post-selection measurement.

\section{Realization in decoherence-free subspaces}

Let us now turn to the question of how to realize our scheme in the decoherence-free subspace $\mathcal{H}=\mathrm{Span}\{\ket{010},\ket{100},\ket{001}\}$.
The decoherence-free subspace not only provides a natural mathematical structure in Eq.~(\ref{eq}), allowing us to use postselection to protect nonadiabatic holonomic operations against the fractional systematic amplitude error, but also gains their resilience to the collective dephasing induced by the interaction Hamiltonian
\begin{align}
H_{\mathrm{int}}=\left[\sigma^{(1)}_{z}+\sigma^{(2)}_{z}+\sigma^{(3)}_{z}\right]\otimes{E},
\end{align}
where $E$ is the environment operator shared by all three qubits \cite{Duan1997,Lidar,Zanardi1997}.
In the three-qubit decoherence-free subspace, the first two qubits are used as the principal subsystem such that $\mathcal{H}_{P}=\mathrm{Span}\{\ket{01},\ket{10},\ket{00}\}$,
and the last qubit is used as the monitoring subsystem such that $\mathcal{H}_{A}=\mathrm{Span}\{\ket{0},\ket{1}\}$.
The computational qubit is encoded as $\ket{0}_{L}\equiv\ket{01}$ and $\ket{1}_{L}\equiv\ket{10}$ while the basis vector $\ket{00}$ acts as an ancilla.
The Hamiltonian governing the time evolution of the quantum system is chosen as
\begin{align}
H(t)=\sum_{k<l}\left[J^{x}_{kl}(t)R^{x}_{kl}+J^{y}_{kl}(t)R^{y}_{kl}\right],
\end{align}
where $J^{x}_{kl}(t)$ and $J^{y}_{kl}(t)$ are the coupling parameters corresponding to the $XY$ interaction  $R^{x}_{kl}=[\sigma^{(k)}_{x}\sigma^{(l)}_{x}+\sigma^{(k)}_{y}\sigma^{(l)}_{y}]/2$ and the Dzyaloshinskii-Moriya interaction $R^{y}_{kl}=[\sigma^{(k)}_{x}\sigma^{(l)}_{y}-\sigma^{(k)}_{y}\sigma^{(l)}_{x}]/2$, respectively \cite{Yang,Johnson,Dzyaloshinsky,Moriya,Li,Wu}. For our purpose, we set the non-zero coupling parameters as $J^{x}_{13}(t)=-J(t)\cos(\theta/2)$, $J^{x}_{23}(t)=J(t)\sin(\theta/2)\cos\varphi$, and
$J^{y}_{13}(t)=J(t)\sin(\theta/2)\sin\varphi$.
Then, we have
\begin{align}
H(t)=J(t)\ket{00}\bra{\Phi_{2}}\otimes\ket{1}\bra{0}+\mathrm{H.c.}
\end{align}
with $\ket{\Phi_{2}}=\sin(\theta/2)\exp(-i\varphi)\ket{0}_{L}-\cos(\theta/2)\ket{1}_{L}$.
Note that the Hamiltonian $H(t)$ has a dark state $\ket{\Phi_{1}}\otimes\ket{0}=\cos(\theta/2)\ket{0}_{L}\ket{0}+\sin(\theta/2)
\exp(i\varphi)\ket{1}_{L}\ket{0}$, where $\ket{\Phi_{1}}$ combines with $\ket{\Phi_{2}}$ making up another basis in the computational space, such that $\mathcal{H}^{a}_{P}(0)
=\mathrm{Span}\{\ket{0}_{L},\ket{1}_{L}\}=\mathrm{Span}\{\ket{\Phi_{1}},\ket{\Phi_{2}}\}$.
If we require $\int^{\tau}_{0}J(t)=\pi$, we have the evolution operator
\begin{align}
\mathcal{U}=\left(\ket{\Phi_{1}}\bra{\Phi_{1}}-\ket{\Phi_{2}}\bra{\Phi_{2}}\right)
\otimes\ket{0}\bra{0}-\ket{00}\bra{00}\otimes\ket{1}\bra{1}.
\end{align}
Recalling that the input states of the principal subsystem are in the computational space spanned by $\{\ket{0}_{L},\ket{1}_{L}\}$, the evolution operator is equivalent to
$\mathcal{U}=(\ket{\Phi_{1}}\bra{\Phi_{1}}-\ket{\Phi_{2}}\bra{\Phi_{2}})\otimes\ket{0}\bra{0}$.
Therefore, the target nonadiabatic holonomic operation can be obtained by tracing out $\ket{0}\bra{0}$ after the time evolution, that is $U=\ket{\Phi_{1}}\bra{\Phi_{1}}-\ket{\Phi_{2}}\bra{\Phi_{2}}$. This is the ideal case without systematic amplitude errors.

If there is a systematic amplitude error $\epsilon$, unlike the ideal case, the evolution operator goes to an erroneous one,
\begin{align}
\mathcal{U}_{\epsilon}=&\left[\ket{\Phi_{1}}\bra{\Phi_{1}}
-\cos(\epsilon\pi)\ket{\Phi_{2}}\bra{\Phi_{2}}\right]\otimes\ket{0}\bra{0}
\notag\\
&-\cos(\epsilon\pi)\ket{00}\bra{00}\otimes\ket{1}\bra{1}
-i\sin(\epsilon\pi)\ket{\Phi_{2}}\bra{00}
\notag\\
&\otimes\ket{0}\bra{1}-i\sin(\epsilon\pi)\ket{00}\bra{\Phi_{2}}\otimes\ket{1}\bra{0}.
\end{align}
This erroneous time evolution operator takes the system initially in the computational space to a state eventually away from the computational subspace.
However, upon performing a conditional measurement on the monitoring subsystem, we can obtain the output state closer to the target output state.
For an input state $\ket{\Psi(0)}=(c_{1}\ket{\Phi_{1}}+c_{2}\ket{\Phi_{2}})\otimes\ket{0}$, the output state is achieved as the following:
\begin{align}
\ket{\Phi_{\epsilon}}=\frac{c_{1}\ket{\Phi_{1}}
-c_{2}\cos(\epsilon\pi)\ket{\Phi_{2}}}{\sqrt{|c_{1}|^{2}+|c_{2}|^{2}\cos^{2}(\epsilon\pi)}},
\end{align}
conditional on $\ket{0}$ being detected.
From the general discussions in the preceding section, we can easily conclude that this output state is much closer to the target output state $\ket{\Phi(\tau)}=c_{1}\ket{\Phi_{1}}-c_{2}\ket{\Phi_{2}}$ than the output state
\begin{align}
\ket{\Phi^{\prime}_{\epsilon}(\tau)}=c_{1}\ket{\Phi_{1}}-c_{2}\cos(\epsilon\pi)\ket{\Phi_{2}}
-ic_{2}\sin(\epsilon\pi)\ket{00}
\end{align}
obtained using the reference scheme.
This ends our discussions on an explicit implementation of our scheme in a decoherence-free subspace.

\section{Conclusion}

In conclusion, we have proposed a scheme to protect nonadiabatic holonomic operations against the systematic amplitude error. Our scheme requires a conditional measurement on a monitor qubit, so that the time-evolving final state can be brought back to the computational subspace when the systematic amplitude error occurs.
Clearly, the monitor qubit introduced in our scheme serves as an error-correction device, through which the impact of systematic amplitude errors on the quantum system is suppressed. In essence, we have thus introduced a measurement-assisted approach to nonadiabatic holonomic operations.
This is markedly different from previous mitigation approaches, namely, the composite pulse nonadiabatic holonomic operations that increase the number of unitary operators \cite{Xu2017} and environment-assisted nonadiabatic holonomic operations that are based on the engineering of an environment \cite{Ramberg}. Our scheme as an alternative may avoid some serious impact of environment-induced decoherence and can work for an actual environment without environment enginnering.
Furthermore, we have given a physical realization of our scheme in a decoherence-free subspace, making it not only robust against the systematic amplitude error but also resilient to some collective dephasing noise.

It is worth noting that our scheme assumes reliable and fast readout of the monitor qubit.
The effectiveness of our scheme will be affected  by the measurement errors and decoherence effects from the monitor qubit.
This is a rather familiar situation.  One similar example is one-way quantum computation, one of the most important models for the realization of quantum computation, implemented by first preparing a highly entangled cluster state and then performing one-qubit measurements on the state \cite{Raussendorf,Raussendorf2003}. Therein the overall computation fidelity will be affected by the state readout errors as well.
Fortunately, current experimental technology permits us to read out a qubit state with high-fidelity and  qubit coherence time can be extended by various active methods.  One trapped-ion qubit is allowed to achieve a single-shot readout fidelity $99.93\%$ \cite{Harty} and its coherence time up to several minutes \cite{Kim}. In superconducting circuits, the transmon qubit admits a measurement fidelity $99.8\%$ \cite{Jeffrey,Yin} and  coherence time up to a dozen microseconds \cite{Barends}.
The feasibility of one-way quantum computation has also been demonstrated experimentally through a universal set of quantum gates, including one-qubit and two-qubit gates \cite{Zeilinger}.
More related to our proposal here, high-fidelity measurement, namely $99.5\%$, has been performed in the experimental demonstration of nonadiabatic holonomic operations \cite{Yin2019}.
Considering that our proposal is a universal scheme independent of specific physical systems, the above-mentioned high-fidelity measurement methods can likely be transferred to the measurement of the ancillary qubit introduced in our scheme.

\begin{acknowledgments}
This work was supported by the National Research Foundation, Singapore and A*STAR under its CQT Bridging Grant.
\end{acknowledgments}


\begin{thebibliography}{99}

\bibitem{Hagley} E. Hagley, X. Maitre, G. Nogues, C. Wunderlich, M. Brune, J. M. Raimond, and S. Hroche, Phys. Rev. Lett. \textbf{79}, 1 (1997).
\bibitem{Turchette} Q. A. Turchette, C. S. Wood, B. E. King, C. J. Myatt, D. Leibfried, W. M. Itano, C. Moroe, and D. J. Wineland, Phys. Rev. Lett. \textbf{81}, 3631 (1998).
\bibitem{Cirac} J. I. Cirac, P. Zoller, H.J. Kimble, and H. Mabuchi, Phys. Rev. Lett. \textbf{78}, 3221 (1997).
\bibitem{Bergmann} K. Bergmann, H. Theuer, and B. W. Shore, Rev. Mod. Phys. \textbf{70}, 1003 (1998).
\bibitem{Bennett} C. H. Bennett, G. Brassard, C. Crepeau, R. Jozsa, A. Peres, and W. K. Wootters, Phys. Rev. Lett. \textbf{70}, 1895 (1993).
\bibitem{Deutsch} D. Deutsch and R. Jozsa, Proc. R. Soc. London A \textbf{439}, 553 (1992).
\bibitem{Chiara} G. De Chiara and G. M. Palma, Phys. Rev. Lett. \textbf{91}, 090404 (2003).
\bibitem{Solonas} P. Solinas, P. Zanardi, and N. Zangh, Phys. Rev. A \textbf{70}, 042316 (2004).
\bibitem{Zhu2005} S. L. Zhu, Z. D. Wang, and P. Zanardi, Phys. Rev. Lett. \textbf{94}, 100502 (2005).
\bibitem{Lipo} C. Lupo, P. Aniello, M. Napolitano, and G. Florio, Phys. Rev. A \textbf{76}, 012309 (2007).
\bibitem{Filipp} S. Filipp, J. Klepp, Y. Hasegawa, C. Plonka-Spehr, U. Schmidt, P. Geltenbort, and H. Rauch, Phys. Rev. Lett. \textbf{102}, 030404 (2009).
\bibitem{Johansson} M. Johansson, E. Sj\"{o}qvist, L. M. Andersson, M. Ericsson, B. Hessmo, K. Singh, and D. M. Tong, Phys. Rev. A \textbf{86}, 062322 (2012).
\bibitem{Berger} S. Berger, M. Pechal, A. A. Abdumalikov, J. C. Eichler, L. Steffen, A. Fedorov, A. Wallraff, and S. Filipp, Phys. Rev. A 87, 060303(R) (2013).
\bibitem{Zanardi} P. Zanardi and M. Rasetti, Phys. Lett. A \textbf{264}, 94 (1999).
\bibitem{Jones} J. A. Jones, V. Vedral, A. Ekert, and G. Castagnoli, Nature {\bf 403}, 869 (2000).
\bibitem{Duan} L. M. Duan, J. I. Cirac, and P. Zoller, Science \textbf{292}, 1695 (2001).
\bibitem{Berry} M. V. Berry, Proc. R. Soc. London, Ser. A {\bf 392}, 45 (1984).
\bibitem{Wilczek} F. Wilczek and A. Zee, Phys. Rev. Lett. {\bf 52}, 2111 (1984).
\bibitem{Messiah} A. Messiah, Quantum Mechanics (North-Holland, Amsterdam, 1962).
\bibitem{Tong} D. M. Tong, Phys. Rev. Lett. \textbf{104}, 120401 (2010).
\bibitem{WangXB} X. B. Wang and K. Matsumoto, Phys. Rev. Lett. {\bf 87}, 097901 (2001).
\bibitem{Zhu}S. L. Zhu and Z. D. Wang, Phys. Rev. Lett. {\bf 89}, 097902 (2002).
\bibitem{Aharonov} Y. Aharonov and J. Anandan, Phys. Rev. Lett. {\bf 58}, 1593 (1987).
\bibitem{Sjoqvist} E. Sj\"{o}qvist, D. M. Tong, L. M. Andersson, B. Hessmo, M. Johansson, and K. Singh, New J. Phys. \textbf{14}, 103035 (2012).
\bibitem{Xu} G. F. Xu, J. Zhang, D. M. Tong, E. Sj\"{o}qvist, and L. C. Kwek, Phys. Rev. Lett. \textbf{109}, 170501 (2012).
\bibitem{Anandan} J. Anandan, Phys. Lett. A \textbf{133}, 171 (1988).
\bibitem{Xu2015} G. F. Xu, C. L. Liu, P. Z. Zhao, and D. M. Tong, Phys. Rev. A \textbf{92}, 052302 (2015).
\bibitem{Sjoqvist2016} E. Sj\"{o}qvist, Phys. Lett. A \textbf{380}, 65 (2016).
\bibitem{Herterich} E. Herterich and E. Sj\"{o}qvist, Phys. Rev. A \textbf{94}, 052310 (2016).
\bibitem{Zhao} P. Z. Zhao, K. Z. Li, G. F. Xu, and D. M. Tong, Phys. Rev. A \textbf{101}, 062306 (2020).
\bibitem{Spiegelberg} J. Spiegelberg and E. Sj\"{o}qvist, Phys. Rev. A \textbf{88}, 054301 (2013).
\bibitem{Liang} Z. T. Liang, Y. X. Du, W. Huang, Z. Y. Xue, and H. Yan, Phys. Rev. A \textbf{89}, 062312 (2014).
\bibitem{Zhang} J. Zhang, L. C. Kwek, E. Sj\"{o}qvist, D. M. Tong, and P. Zanardi, Phys. Rev. A \textbf{89}, 042302 (2014).
\bibitem{Xue} Z. Y. Xue, J. Zhou, and Z. D. Wang, Phys. Rev. A \textbf{92}, 022320 (2015).
\bibitem{Xue2016} Z. Y. Xue, J. Zhou, Y. M. Chu, and Y. Hu, Phys. Rev. A \textbf{94}, 022331 (2016).
\bibitem{Xue2017} Z. Y. Xue, F. L. Gu, Z. P. Hong, Z. H. Yang, D. W. Zhang, Y. Hu, and J. Q. You, Phys. Rev. Appl. \textbf{7}, 054022 (2017).
\bibitem{Zhao2017} P. Z. Zhao, X. D. Cui, G. F. Xu, E. Sj\"{o}qvist, and D. M. Tong, Phys. Rev. A \textbf{96}, 052316 (2017).
\bibitem{Hong} Z. P. Hong, B. J. Liu, J. Q. Cai, X. D. Zhang, Y. Hu, Z. D. Wang, and Z. Y. Xue, Phys. Rev. A \textbf{97}, 022332 (2018).
\bibitem{Xue2018} T. Chen, J. Zhang, and Z. Y. Xue, Phys. Rev. A \textbf{98}, 052314 (2018).
\bibitem{Zhao2019} P. Z. Zhao, G. F. Xu, and D. M. Tong, Phys. Rev. A \textbf{99}, 052309 (2019).
\bibitem{Xue2020} T. Chen, P. Shen, and Z. Y. Xue, Phys. Rev. Appl. \textbf{14}, 034038 (2020).
\bibitem{Xue2021} S. Li and Z. Y. Xue, Phys. Rev. Appl. \textbf{16}, 044005 (2021).
\bibitem{Xue2022} Y. Liang, P. Shen, T. Chen, and Z. Y. Xue, Phys. Rev. Appl. \textbf{17}, 034015 (2022).
\bibitem{Zhao2023} P. Z. Zhao and D. M. Tong, Phys. Rev. A \textbf{108}, 012619 (2023).
\bibitem{Liu} B. J. Liu, X. K. Song, Z. Y. Xue, X. Wang, and M. H. Yung, Phys. Rev. Lett. \textbf{123}, 100501 (2019).
\bibitem{Zhang2023} J. Zhang, T. H. Kyaw, S. Filipp, L. C. Kwek, E. Sj\"{o}qvist, D. M. Tong, Phys. Rep. \textbf{1027}, 1 (2023).
\bibitem{Abdumalikov} A. A. Abdumalikov, J. M. Fink, K. Juliusson, M. Pechal, S. Berger, A. Wallraff, and S. Filipp, Nature \textbf{496}, 482 (2013).
\bibitem{Long} G. R. Feng, G. F. Xu, and G. L. Long, Phys. Rev. Lett. \textbf{110}, 190501 (2013).
\bibitem{Duan2014} C. Zu, W. B. Wang, L. He, W. G. Zhang, C. Y. Dai, F. Wang, and L. M. Duan, Nature \textbf{514}, 72 (2014).
\bibitem{Camejo} S. A. Camejo, A. Lazariev, S. W. Hell, and G. Balasubramanian, Nat. Commun. \textbf{5}, 4870 (2014).
\bibitem{Sekiguchi} Y. Sekiguchi, N. Niikura, R. Kuroiwa, H. Kano, and H. Kosaka, Nat. Photonics \textbf{11}, 309 (2017).
\bibitem{Sun} Y. Xu, W. Cai, Y. Ma, X. Mu, L. Hu, Tao Chen, H. Wang, Y. P. Song, Z. Y. Xue, Z. Q. Yin, and L. Sun, Phys. Rev. Lett. \textbf{121}, 110501 (2018).
\bibitem{Yin2019} Z. X. Zhang, P. Z. Zhao, T. H. Wang, L. Xiang,Z. L. Jia, P. Duan, D. M. Tong, Y. Yin, and G. P. Guo, New J. Phys. \textbf{21}, 073024 (2019).
\bibitem{Zhou} B. B. Zhou, P. C. Jerger, V. O. Shkolnikov, F. J. Heremans, G. Burkard, and D. D. Awschalom, Phys. Rev. Lett. \textbf{119}, 140503 (2017).
\bibitem{Nagata} K. Nagata, K. Kuramitani, Y. Sekiguchi, and H. Kosaka, Nat. Commun. \textbf{9}, 3227 (2018).
\bibitem{Jin} J. Jing, C. H. Lam, and L. A. Wu, Phys. Rev. A \textbf{95}, 012334 (2017).
\bibitem{Jin2024} Z. Y. Jin and J. Jing, Phys. Rev. A \textbf{109}, 012619 (2024).
\bibitem{Vandersypen} L. M. K. Vandersypen and I. L. Chuang, Rev. Mod. Phys. \textbf{76}, 1037 (2005).
\bibitem{Monz} T. Monz, K. Kim, W. H\"{a}nsel, M. Riebe, A. S. Villar, P. Schindler, M. Chwalla, M. Hennrich, and R. Blatt, Phys. Rev. Lett. \textbf{102}, 040501 (2009).
\bibitem{Timoney} N. Timoney, V. Elman, S. Glaser, C. Weiss, M. Johanning, W. Neuhauser, and Chr. Wunderlich, Phys. Rev. A \textbf{77}, 052334 (2008).
\bibitem{Xu2017} G. F. Xu, P. Z. Zhao, T. H. Xing, E. Sj\"{o}qvist, and D. M. Tong, Phys. Rev. A \textbf{95}, 032311 (2017).
\bibitem{Ramberg} N. Ramberg and E. Sj\"{o}qvist, Phys. Rev. Lett. \textbf{122}, 140501 (2019).
\bibitem{Pellizzari} T. Pellizzari, S. A. Gardiner, J. I. Cirac, and P. Zoller, Phys. Rev. Lett. \textbf{75}, 3788 (1995).
\bibitem{Cirac1995} J. I. Cirac and P. Zoller, Phys. Rev. Lett. \textbf{74}, 4091 (1995).
\bibitem{Blais} A. Blais, R. S Huang, A. Wallraff, S. M. Girvin, and R. J. Schoelkopf, Phys. Rev. A \textbf{69}, 062320 (2004).
\bibitem{Blais2007} A. Blais, J. Gambetta, A. Wallraff, D. I. Schuster, S. M. Girvin, M. H. Devoret, and R. J. Schoelkopf, Phys. Rev. A \textbf{75}, 032329 (2007).
\bibitem{Duan1997} L. M. Duan and G. C. Guo, Phys. Rev. Lett. \textbf{79}, 1953 (1997).
\bibitem{Zanardi1997} P. Zanardi and M. Rasetti, Phys. Rev. Lett. \textbf{79}, 3306 (1997).
\bibitem{Lidar} D. A. Lidar, I. L. Chuang, and K. B. Whaley, Phys. Rev. Lett. \textbf{81}, 2594 (1998).
\bibitem{Yang} C. N. Yang and C. P. Yang, Phys. Rev. \textbf{150}, 321 (1966).
\bibitem{Johnson} J. D. Johnson and M. McCoy, Phys. Rev. A \textbf{6}, 1613 (1972).
\bibitem{Dzyaloshinsky} L. Dzyaloshinsky, J. Phys. Chem. Solids \textbf{4}, 241 (1958).
\bibitem{Moriya} T. Moriya, Phys. Rev. Lett. \textbf{4}, 228 (1960).
\bibitem{Li} K. Z. Li, P. Z. Zhao, and D. M. Tong, Phys. Rev. Research \textbf{2}, 023295 (2020).
\bibitem{Wu} X. Wu and P. Z. Zhao, Phys. Rev. A \textbf{102}, 032627 (2020).
\bibitem{Raussendorf} R. Raussendorf and H. J. Briegel, Phys. Rev. Lett. \textbf{86}, 5118 (2001).
\bibitem{Raussendorf2003} R. Raussendorf, D. E. Browne, and H. J. Briegel, Phys. Rev. A \textbf{68}, 022312 (2003).
\bibitem{Harty} T. P. Harty, D. T. C. Allcock, C. J. Ballance, L. Guidoni, H. A. Janacek, N. M. Linke, D.N. Stacey, and D. M. Lucas, Phys. Rev. Lett. \textbf{113}, 220501 (2014).
\bibitem{Kim} Y. Wang, M. Um, J. Zhang, S. An, M. Lyu, J. N. Zhang, L. M. Duan, D. Yum and K. Kim, Nat. Photon. \textbf{11}, 646 (2017).
\bibitem{Jeffrey} E. Jeffrey, D. Sank, J. Y. Mutus, T. C. White, J. Kelly, R. Barends, Y. Chen, Z. Chen, B. Chiaro, A. Dunsworth, A. Megrant, P. J. J. O'Malley, C. Neill, P. Roushan, A. Vainsencher, J. Wenner, A. N. Cleland, and J. M. Martinis, Phys. Rev. Lett. \textbf{112}, 190504 (2014).
\bibitem{Yin} T. H. Wang, Z. X. Zhang, L. Xiang, Z. L. Jia, P. Duan, W. Z. Cai, Z. H. Gong, Z. W. Zong, M. M. Wu, J. L. Wu, L. Y. Sun, Y. Yin, and G. P. Guo, New J. Phys. \textbf{20}, 065003 (2018).
\bibitem{Barends} R. Barends, J. Kelly, A. Megrant, A. Veitia, D. Sank, E.Jeffrey, T. C. White, J. Mutus, A. G. Fowler, B. Campbell, Y. Chen, Z. Chen, B. Chiaro, A. Dunsworth, C. Neill, P. O'Malley, P. Roushan, A. Vainsencher, J. Wenner, A. N. Korotkov, A. N. Cleland, and J. M. Martinis, Nature \textbf{508}, 500 (2014).
\bibitem{Zeilinger} P. Walther, K. J. Resch, T. Rudolph, E. Schenck, H. Weinfurter, V. Vedral, M. Aspelmeyer, and A. Zeilinger, Nature \textbf{434}, 169 (2006).

\end{thebibliography}
\end{document}